\documentclass[russian]{article}
\usepackage[T1]{fontenc}
\usepackage[koi8-r]{inputenc}
\usepackage[letterpaper]{geometry}
\usepackage{textcomp}
\usepackage{graphicx}
\usepackage{setspace}

\makeatletter

\AtBeginDocument{\DeclareFontEncoding{T2A}{}{}}

\newcommand{\lyxaddress}[1]{
\par {\raggedright #1
\vspace{1.4em}
\noindent\par}
}

\makeatother

\usepackage{babel}

\begin{document}

\title{Crystallization of amorphous alloy of Fe-Cu-Nb-Si-B under the influence
of high-power flashing optical radiation}

\author{Nazipov R.A., Zuzin N.A., Mitin A.V.$^*$}

\maketitle

\lyxaddress{\textit{Physics Faculty, Kazan (Volga region) Federal University,
18 Kremlyovskaya St., Kazan 420008, Russia}}

\lyxaddress{\textit{Казанский (Приволжский) федеральный университет, 420008,
Казань, Кремлевская, 18, физический факультет}}

\lyxaddress{\textit{$^*$Institute of polymers, Kazan State Technological University,
72 {}``B'' Karl Marx St., Kazan 420015, Russia}}

\lyxaddress{\textit{Казанский государственный технологический университет, 420015,
Казань, ул. Карла Маркса, д.72, Корп. \textquotedbl{}Б\textquotedbl{},
Институт полимеров}}
\begin{abstract}
\noindent
Investigation of crystallization in an 5BDSR amorphous alloy
by system Fe-Cu-Nb-Si-B under the influence of an incoherent optical
radiation generated by gas discharge flash-lamp using X-ray diffraction
scattering. It is shown that depending on the magnitude of the input
power to the flash-lamp of structure of the alloy varies widely after
radiation: a change in the short-range order during retaining the
amorphous structure or the formation of several crystalline phases
with different quantitative content in the alloy. Conditions are found
for the formation of the nanocrystalline structure. Analysis in terms
of the Kolmogorov-Johnson-Mehl-Avrami kinetics showed that the mechanism
of primary crystallization under flash irradiation is associated with
a well-known for finemet-type alloys: rapid nucleation, followed by
growth slowing of nanocrystalline phase $\alpha$-Fe(Si) with the
D0$_3$ structure.

-----

Методами рентгеновской дифракции исследовалась кристаллизация аморфного
сплава 5БДСР системы Fe-Cu-Nb-Si-B под действием оптического некогерентного
излучения, генерируемого газоразрядной лампой-вспышкой. Показано,
что в зависимости от величины подведенной энергии к лампе-вспышки
структура сплава после облучения меняется в широких пределах: от изменения
ближнего порядка, с сохранением аморфной структуры, до кристаллизации,
с образованием нескольких фаз, с разным количественным содержанием
их в сплаве. Найдены условия образования нанокристаллической структуры.
На основе анализа кинетики кристаллизации нанокристаллической фазы,
с привлечением модели Колмогорова-Джонсона-Мела-Аврами, установлено,
что механизм первичной кристаллизации связан с хорошо известным для
сплавов finemet-типа быстрым зарождением центров кристаллизации с
последующим замедлением роста нанокристаллов фазы $\alpha$-Fe(Si)
с D0$_{3}$ структурой.

\pagebreak{}
\end{abstract}
\begin{center}
{\large Кристаллизация аморфного сплава системы Fe-Cu-Nb-Si-B под
действием мощного импульсного оптического излучения}
\par\end{center}{\large \par}

\begin{onehalfspace}
\begin{center}
Назипов Р.А., Зюзин Н.А., Митин А.В.
\par\end{center}
\end{onehalfspace}

\section{Введение}

Поиск новых способов термической обработки аморфных сплавов, которые
смогли бы упростить существующую технологию и в то же время улучшить
их функциональные свойства, является актуальным направлением физического
материаловедения. Магнитные сплавы finemet-типа, на основе композиции
Fe-Cu-Nb-Si-B с преимущественным содержанием железа (около 74-77\%),
обладают особенностью кристаллизоваться из аморфного состояния в композитную
структуру с беспорядочно ориентированными нанокристаллами в аморфной
матрице \cite{FINE001}. Размер нанокристаллов составляет около 10~нм,
что намного меньше, чем длина блоховской стенки. Это приводит к тому,
что в объеме кристаллографическая анизотропия отдельного зерна усредняется
и среднее значение анизотропии стохастического магнитного домена становится
очень малым. Как следствие, такие сплавы после термической обработки
обладают очень низкой коэрцитивной силой, с одновременно довольно
высокими значениями индукции насыщения и магнитной проницаемости \cite{NANO001,patent5591276}.

Стандартная технология, при которой формируется наноструктура, заключается
в отжиге сформированных из сплава Fe-Cu-Nb-Si-B заготовок в печи в
течение 30-60~мин. при температуре около 550$^\circ$С. В результате
отжига сплав не только приобретает уникальные магнитные свойства,
но также становится очень хрупким и поверхность его окисляется. Для
предотвращения окисления используют термический отжиг в атмосфере
инертного газа или в вакууме. Большая хрупкость сплава после отжига
является значительной проблемой. Это обуславливает необходимость изготавливать
заготовку заранее и отжигать её целиком. С целью уменьшения окисления
поверхности аморфного сплава и улучшения механических свойств, с одновременным
улучшением магнитных свойств были предприняты попытки использовать
быстрый отжиг аморфных сплавов \cite{FLASH006,FLASH001,FLASH002}.
Однако такая обработка применялась в смысле отпуска избыточных напряжений,
а не формирования нанокристаллической структуры. Для создания равномерной
нанокристаллической структуры необходимо создать условия, при которых
по всему объему однородно формируются множество центров кристаллизации,
и затем важно ограничить рост кристаллитов до величины около 10~нм.
В сплавах состава Fe-Si-B этого не происходит \textemdash{} кристаллические
зерна вырастают до значительных размеров и при этом сильно разнятся
форма и величина зерен. Появление множества центров кристаллизации
можно добится добавлением в Fe-Si-B порядка 1~ат.~\% Cu. В этом
случае в сплаве Fe-Si-B происходит сегрегация меди по всему объему
в виде кластеров с ГЦК структурой размером 3-5~нм \cite{HONO007}.
Важно отметить, что температура при которой происходит выделение медных
кластеров гораздо ниже, чем температура образования первичной кристаллической
фазы. На границе медных кластеров сильно понижается энергия активации
кристаллизационных зародышей фазы $\alpha$-Fe(Si). Но они все еще
могут вырасти до значительных размеров, поэтому к сплаву добавляют
около 3~ат.~\% Nb, что ограничивает рост фазы $\alpha$-Fe(Si) и
приводит к формированию нанокристаллической структуры в довольно широком
температурном диапазоне (550-650$^\circ$С).

На процесс кристаллизации значительное влияние оказывает скорость
нагрева аморфного сплава. Увеличение скорости нагрева аморфных сплавов
приводит к увеличению скорости кристаллизации \cite{KINET001}, при
этом температуры кристаллизации первой и второй стадий смещаются в
большую сторону. Смещение температур кристаллизации для первой и второй
стадий происходит неравномерно и для высоких скоростей нагрева нет
данных калориметрии о том, насколько близко или далеко друг от друга
на температурной шкале находятся максимумы скоростей кристаллизации.
Возможно при высокой скорости нагрева они находяться ближе друг к
другу, что может затруднить получение сплава системы Fe-Cu-Nb-Si-B
в нанокристаллическом состоянии при импульсном отжиге. Кроме того,
до конца не было ясно, успеют ли атомы меди сгруппироваться в кластеры
во время импульсного отжига и при какой подводимой энергии это произойдет.

Ранее в работе \cite{NAZ2010} сообщалось о исследовании кристаллизации
в сплаве 5БДСР после импульсного отжига излучением мощного электрического
разряда в атмосфере воздуха. Электрический разряд в воздухе \textemdash{}
сложный процесс, при котором помимо электромагнитного излучения на
образец действует ударная волна быстро разогретого воздуха. Существенное
влияние на процесс отжига этим методом оказывает флюктуация подводимой
к образцу энергии, вследствие стохастической природы импульсного электрического
разряда, поэтому довольно трудно получить промежуточное аморфное-кристаллическое
структурное состояние. Для уменьшения флюктуации подводимой к образцу
энергии, в качестве источника мощного импульсного электромагнитного
излучения в диапазоне от инфракрасного (ИК) до ультрафиолетового (УФ)
мы использовали лампу-вспышку \textemdash{} газоразрядную лампу с
внешним поджигом. В работе \cite{NAZFLASH} была показана возможность
кристаллизации аморфного сплава на основе Fe под действием мощного
импульсного некогерентного электромагнитного излучения, генерируемого
лампой-вспышкой типа ИФК-2000. При этом исключается действие воздушной
ударной волны и влияние плазмы электрического разряда, что позволяет
поместить образец вплотную к колбе лампы. Для возможного практического
применения интерес представляет отсутствие окислов на поверхности
ленты при кристаллизации под действием импульсного электромагнитного
излучения. Однако, при подводимой энергии, значительно большей чем
необходимо для кристаллизации образца по объему, происходит горение
сплава с образованием сложных окислов. Протекающий в лампе-вспышке
импульсный разряд довольно стабилен, и величина излучаемой энергии
электромагнитного излучения является хорошо предсказуемой. Поэтому
влияние импульсного излучения на структуру аморфного сплава в области
структурного перехода аморфный-кристаллический лучше повторяется,
чем при облучении электрическим разрядом в воздухе.

В настоящей работе мы показываем принципиальную возможность получения
сплава системы Fe-Cu-Nb-Si-B в нанокристаллическом состоянии при импульсном
отжиге аморфного сплава некогерентным электромагнитным излучением
в диапазоне от инфракрасного до ультрафиолетового.

\section{Эксперимент}

В качестве объекта исследований был выбран аморфный сплав 5БДСР производства
Ашинского металлургического завода (5БДСР тип В, ТУ 14-123-149-99,
номер плавки 2-5-759), который относится к сплавам системы Fe-Cu-Nb-Si-B.
Сплав был изготовлен методом спиннингования и представляет собой тонкую
ленту, толщиной около 25~мкм и шириной 10~мм. Из ленты были вырезаны
образцы, длиной 30~мм, которые и подвергались отжигу в атмосфере
воздуха излучением лампы-вспышки XV80 производства чехословацкой компании
TESLA.

Колба лампы-вспышки XV80 имеет соленоидальный вид, с внутренним диаметром
15~мм, на рис. 1 приведена её фотография.

\begin{figure}
\begin{centering}
\includegraphics[scale=0.1]{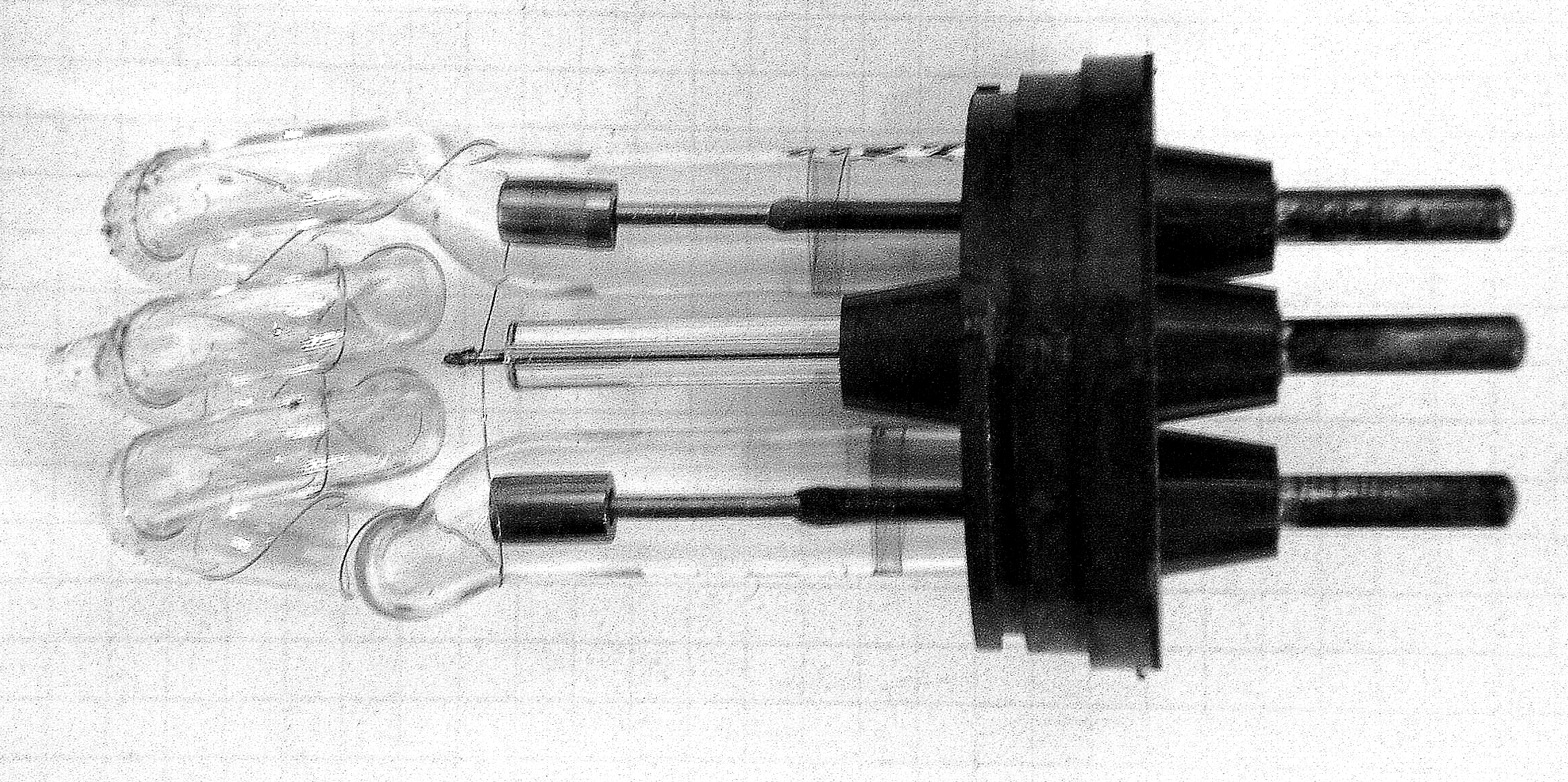}
\par\end{centering}

\caption{Внешний вид лампы-вспышки XV80 производства компании TESLA (Чехословакия,
изготовлена в 1983~г.)}

\end{figure}

Образец помещался внутри лампы на оси колбы-катушки таким образом,
чтобы центр тяжести образца был примерно посередине лампы.

В качестве источника электрической энергии лампы использовалась установка,
включающая в себя:
\begin{itemize}
\item высоковольтный блок питания постоянного напряжения на базе УЭЛИ-1;
\item конденсаторную батарею общей емкостью 1179~мкФ, собранную из 8-ми
конденсаторов ИМ5-150, подключенных параллельно;
\item устройство коммутации, состоящей из измерительного устройства высокого
напряжения, высоковольтного трансформатора Тесла для поджига разряда
в лампе, ключа заряда и ЛАТРа.
\end{itemize}
Принципиальная электрическая схема установки приведена на рис.~2.

\begin{figure}
\begin{centering}
\includegraphics[scale=0.13]{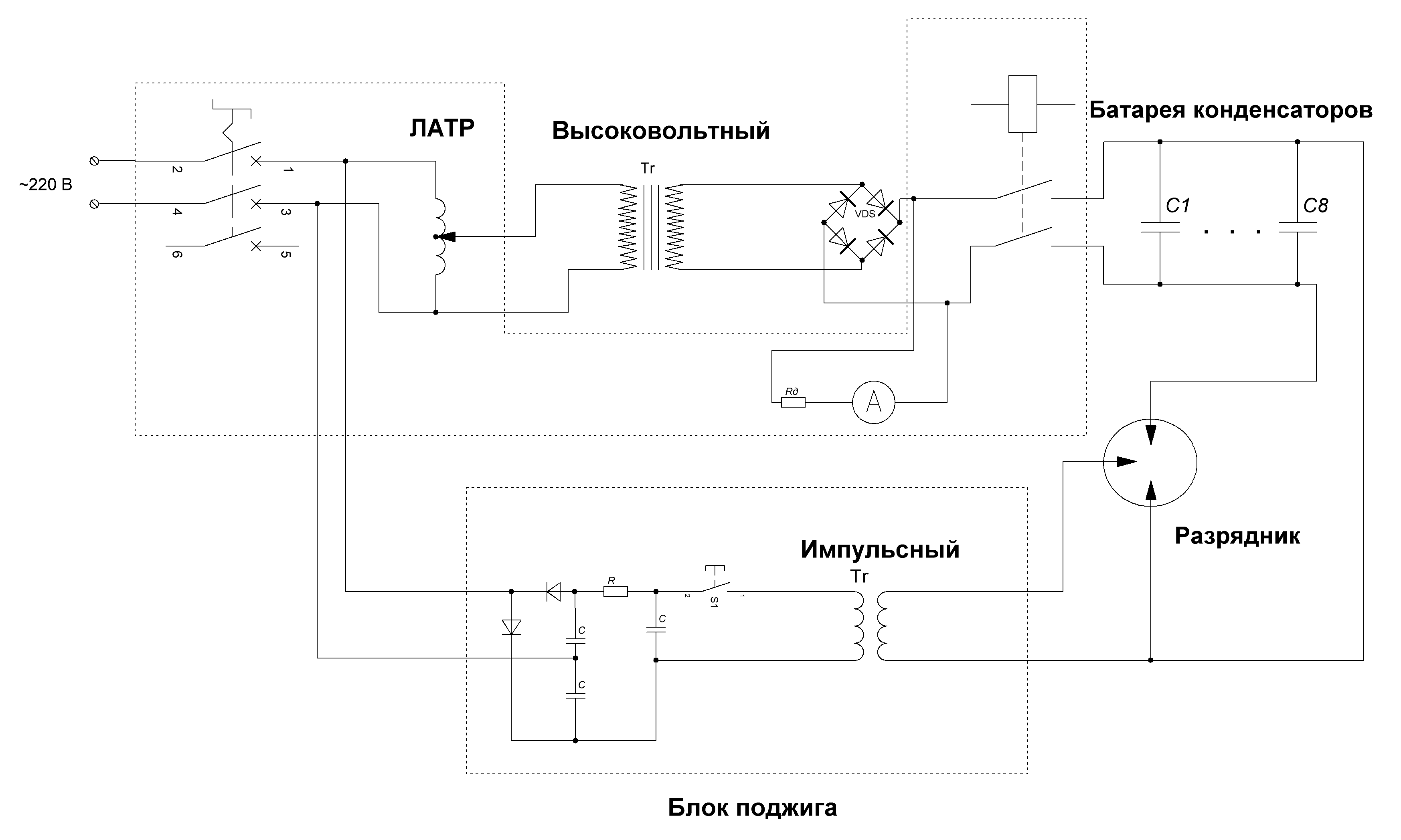}
\par\end{centering}

\caption{Принципиальная схема установки для питания лампы или разрядника и
системы поджига}

\end{figure}

Длительность электромагнитного импульса составляет порядка 500~мкс.
Для оценки подводимой к лампе электрической энергии использовалось
соотношение $E_{e}=CU^{2}/2$.

Импульсное облучение образцов сплава 5БДСР проводилось в несколько
серий:
\begin{enumerate}
\item Облучение одним импульсом, с разной подводимой электрической энергией
\textemdash{} 747~Дж, 917~Дж, 1103~Дж, 1307~Дж и 1529~Дж (нецелые
числа энергий связаны с особенностью измерения высокого напряжения).
\item Облучение при одинаковой подводимой электрической энергии \textit{E}$_e$~=~1307~Дж,
но с разным количеством импульсов: \textit{N}~=~1, 2, 5, 10, 20
и 30.
\item Облучение при одинаковой подводимой электрической энергии \textit{E}$_e$~=~917~Дж,
но с разным количеством импульсов: \textit{N}~=~1, 5, 10, 20, 30,
40 и 50.
\end{enumerate}
Отдельно были облучены 20-ю импульсами образцы с подводимой энергией
\textit{E}$_e$~=~747~Дж (\textit{N}~=~20), а также с подводимой
энергией \textit{E}$_e$=1103~Дж (\textit{N}~=~20).

Рентгеноструктурные исследования проводились на рентгеновском минидифрактометре
МД-10 \guillemotleft{}ЭФА\guillemotright{} (Производства ЗАО \guillemotleft{}Радикон\guillemotright{},
г.~С.-Петербург), с регистрацией излучения позиционно-чувствительным
детектором (ПЧД). В качестве источника рентгеновского излучения применялась
трубка прострельного типа, маломощная, охлаждаемая потоком воздуха,
с железным анодом. Излучение Fe-$K_{\alpha}$ было выделено кристаллом-монохроматором
LiF. Следует указать на оригинальную рентгенооптическую схему минидифрактометра,
вся область углов 2$\theta$ разбита на два диапазона: малых углов
и больших углов. Поскольку ПЧД неподвижен во все время измерений,
диапазон меняется изменением угла ввода первичного рентгеновского
пучка. В первом диапазоне угол ввода первичного пучка с плоскостью
образца составляет около 5$^\circ$, что позволяет исследовать поверхностный
слой на глубине нескольких мкм. Однако в этой геометрии небольшие
смещения по вертикальной оси плоскости образца приводят к заметному
смещению дифракционных рефлексов, что не позволяет использовать прибор
для прецизионного определения межплоскостных расстояний. Тем не менее,
при фазовом анализе, когда необходимо измерить относительные интенсивности
рефлексов, данный прибор позволяет получить очень точные результаты
в пределах одного диапазона 2$\theta$. Рентгеноструктурные исследования
проводились во всех случаях с матовой (неконтактной) стороны ленты.

\section{Результаты и обсуждение}

После облучения лампой-вспышкой импульсным некогерентным оптическим
излучением, все образцы сплава 5БДСР оставались неокисленными, практически
без следов побежалости. Поверхность ленты становилась несколько неровной
и становилась хрупкой, однако не катастрофически \textemdash{} допускала
небольшой изгиб до излома даже при полной кристаллизации.

На рис.~3 приведены рентгеновские дифрактограммы, полученные до и
после облучения сплава 5БДСР с разной подводимой энергии за один импульс.
Из этого рисунка видно, что в исходном состоянии лента является аморфной.
По мере увеличения подводимой энергии форма профиля рентгеноаморфного
гало становится сначала асимметричной (\textit{E}$_e$~=~917~Дж),
а затем сужается (\textit{E}$_e$~=~1103~Дж). Кристаллизация сплава
происходит при облучении с подводимой энергией 1307~Дж и 1529~Дж,
в последнем случае вид дифрактограммы такой же, как при термическом
отжиге ленты 5БДСР при 900$^\circ$С. Поэтому мы пришли к выводу,
что при облучении сплава 5БДСР с подводимой энергией 1529~Дж происходит
полная кристаллизация образца. Здесь интересно отметить, что обычный
термический отжиг аморфной металлической ленты до полной кристаллизации
следует проводить в вакууме или в инертной атмосфере, чтобы избежать
её окисления.

\begin{figure}
\begin{centering}
\includegraphics[width=10cm]{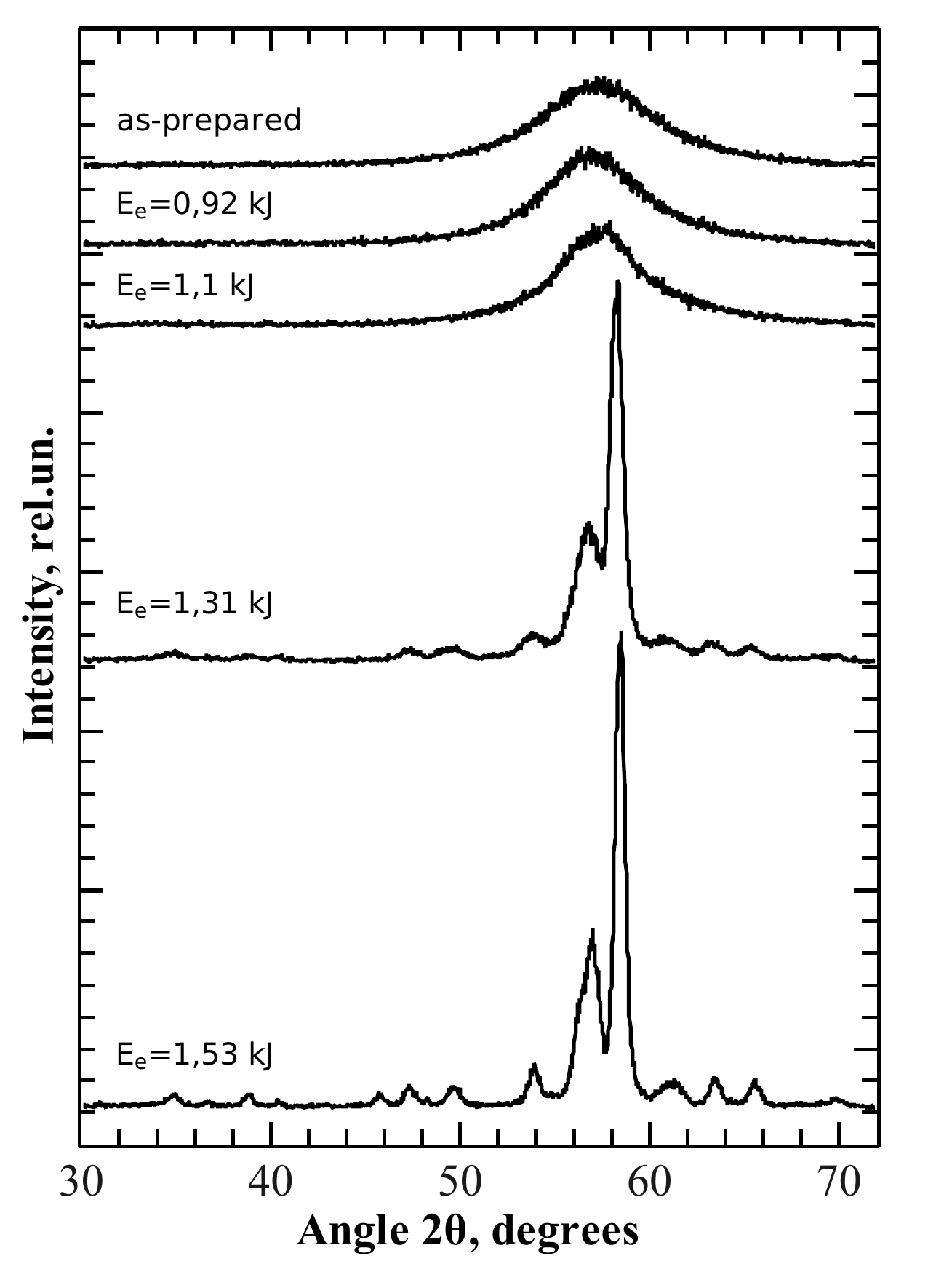}
\par\end{centering}

\caption{Дифрактограммы сплава 5БДСР в исходном состоянии и после облучения
одиночным импульсом некогерентного оптического излучения с разной
подводимой энергией}

\end{figure}

Фазовый состав сплава 5БДСР после полной кристаллизации как минимум
двухкомпонентный и включает в себя фазу $\alpha$-Fe(Si) со структурой
неупорядоченной сверхрешетки D0$_3$ и гексагональную фазу, которая
обозначается как H-фаза \cite{LAS001,LAS002}. На рис.~4 приведена
дифрактограмма полностью кристаллизованного образца, полученного облучением
аморфного сплава одиночным импульсом с \textit{E}$_e$~=~1529~Дж.
Дифрактограмма построена по логарифмической шкале межплоскостных расстояний
\textit{d}(\r{A}), на ней рядом с основными рефлексами указаны фазы
и их индексы отражающих плоскостей (\textit{hkl}).

\begin{figure}
\begin{centering}
\includegraphics[scale=0.3]{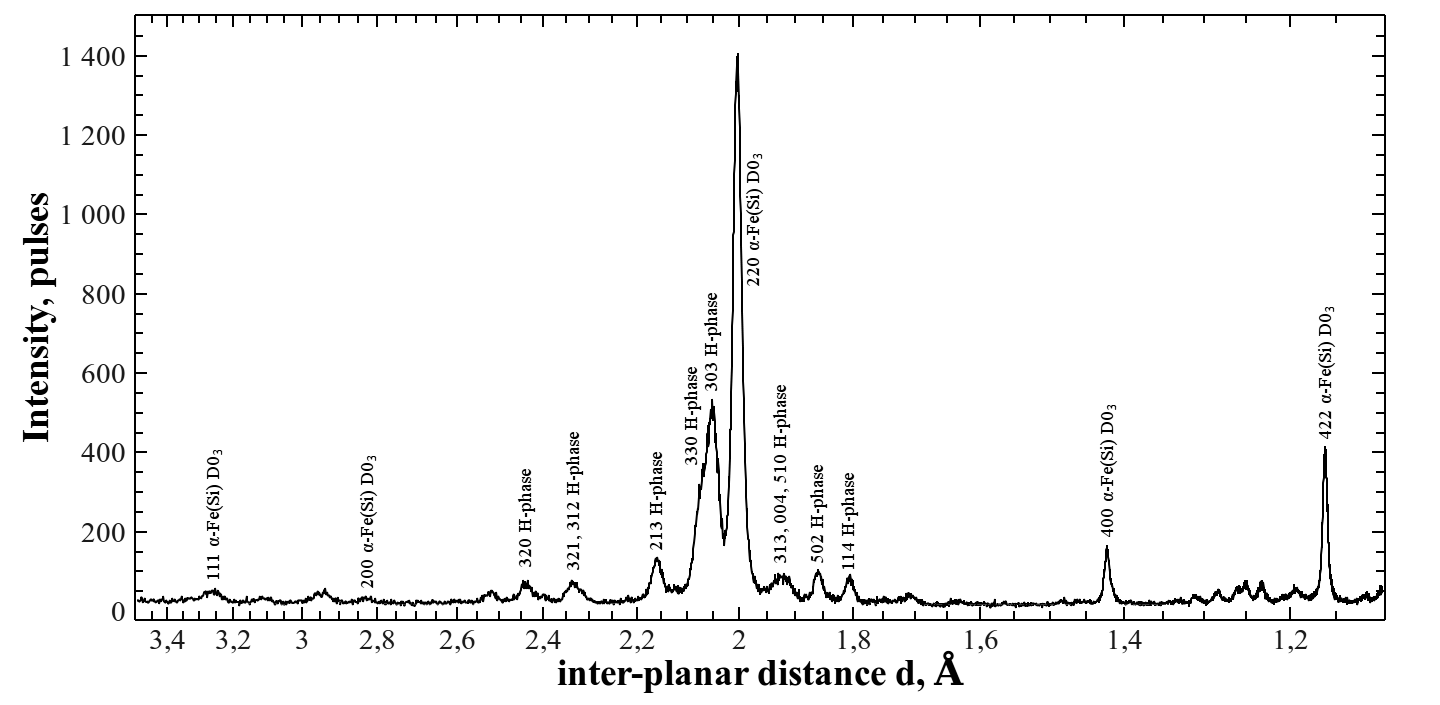}
\par\end{centering}

\caption{Дифрактограмма полностью кристаллизованного образца}

\end{figure}

Дифрактограмма образца, облученного с \textit{E}$_e$~=~747~Дж,
даже при увеличении количества импульсов до \textit{N}~=~20 остается
практически без изменений. Поскольку заметные изменения дифракционного
профиля рентгеноаморфного гало наблюдаются для образцов, облученных
с \textit{E}$_e$~=~917~Дж и \textit{E}$_e$~=~1103~Дж, было
сделано предположение, что увеличение количества импульсов, при данной
неизменной подводимой энергии, может вызвать большие изменения в структуре
сплава. Оказалось, что при облучении 20-ю импульсами с \textit{E}$_e$~=~917~Дж,
образуется нанокристаллическая фаза $\alpha$-Fe(Si), со средним размером
кристаллитов около 11\textpm{}3~нм, широкий рефлекс которой появляется
на фоне остаточной аморфной фазы. Отжиг 20-ю импульсами с подводимой
энергией \textit{E}$_e$~=~1103~Дж приводит к формированию H-фазы,
одновременно с нанокристаллической фазой $\alpha$-Fe(Si). Причем
интересно отметить, что рефлекс от H-фазы трансформируется из дифракционного
гало и имеет сильно уширенную форму, вероятно, вследствие дисперсности.

Для анализа кинетики кристаллизации в материалах часто используют
модель Колмогорова-Джонсона-Мела-Аврами (КДМА) \cite{KOL001}. Эта
модель основана на представлении процесса кристаллизации, состоящего
из двух этапов: появлении случайным образом кристаллизационных зародышей
в некотором объеме, и рост кристаллической фазы из этих зародышей
с некоторой скоростью и до некоторой величины. Строго говоря, эта
модель справедлива для изотермического отжига и выражает зависимость
объемной доли \textit{x} выделяющейся кристаллической фазы от времени
\textit{t} \cite{CRYS003}:

\begin{equation}
x(t)=1-\exp\left[-\left(Kt\right)^{n}\right],\end{equation}

где \textit{n} \textemdash{} показатель Аврами или кинетический показатель,
отвечающий за размерность растущих кристаллов и частоту появления
кристаллических зародышей, а \textit{K} \textemdash{} коэффициент
Аврами или постоянная скорости реакции (кристаллизации). Показатель
Аврами состоит из двух слагаемых $n=n_{n}+n_{g}$, где $n_{n}$ \textemdash{}
отвечает за появление кристаллических зародышей, а $n_{g}$ \textemdash{}
за рост кристаллической фазы. В случае 3-х мерного роста кристаллических
зерен nn имеет значения $0\leq n_{n}\leq1$: $n_{n}=0$ \textemdash{}
кристаллические зародыши уже существовали ранее и их количество было
постоянно либо скорость их появления сильно замедленна; $n_{n}=1$
\textemdash{} скорость зарождения кристаллических зародышей постоянна,
т.е. они появляются одновременно с ростом кристаллических зерен.

В свою очередь $n_{g}$ имеет значения от 3 при линейном росте кристаллитов,
до 1,5 при параболическом или диффузионно-барьерном росте. Постоянная
скорости реакции \textit{K} зависит от температуры $T$ по закону
Аррениуса:

\begin{equation}
K(T)=K_{0}\exp\left(\frac{-E_{a}}{kT}\right),\end{equation}

где \textit{$K_{0}$} \textemdash{} кинетический параметр постоянной
скорости появления кристаллических зародышей, по порядку величины
он близок к частоте колебаний атомов около положения равновесия ($K_{0}\approx10^{13}$~Гц);
$E_{a}$ \textemdash{} энергия активации для появления зародыша кристаллической
фазы.

Значения \textit{n} и \textit{K} можно найти экспериментально. Для
этого необходимо дважды прологарифмировать уравнение (1) для получения
следующего соотношения:

\begin{equation}
\ln\left[-\ln\left(1-x(t)\right)\right]=\ln K+n\cdot\ln t\,.\end{equation}

Найдя экспериментальные значения \textit{x}(\textit{t}), строят по
точкам график $\ln\left[-\ln\left(1-x(t)\right)\right]$ от $\ln t$
и затем, апроксимируя экспериментальные точки линией, находят значения
\textit{n} и ln\textit{K}. Далее по значению \textit{n} делают предположения
о преобладающем механизме кристаллизации и находят энергию активации
$E_{a}$.

На рис.~5 представлены дифрактограммы сплава после облучения разным
количеством импульсов при подводимой энергии $E_{e}$~=~917~Дж.
Эти дифрактограммы явно демонстрируют уменьшение относительной объемной
доли аморфной фазы и увеличение нанокристаллической фазы при увеличении
количества импульсов. Для нахождения относительного содержания нанокристаллической
фазы \textit{x}(\textit{t}) из рентгенодифракционных экспериментальных
данных используется связь между интегральной интенсивностью рефлекса
анализируемой фазы и её объемной долей:

\begin{equation}
x(t)=\frac{I_{c}}{I_{c}+I_{a}}\,,\end{equation}

где $I_{c}$ \textemdash{} интегральная интенсивность нанокристаллической
фазы, а $I_{a}$ \textemdash{} интегральная интенсивность аморфной
фазы. В программе Fityk \cite{FIT001} были найдены параметры профилей
(положение центра тяжести, амплитуда, интегральная интенсивность линий
и т.п.) от аморфного дифракционного гало и дифракционных рефлексов,
принадлежащих нанокристаллической фазе. В качестве анализируемой функции
дифракционных профилей была выбрана функция Пирсона VII. При использовании
этой функции достигалось наилучшее согласие между экспериментальными
и вычисленными дифрактограммами. Фон был аппроксимирован линейной
функцией. Из найденных значений интегральных интенсивностей рефлексов
нанокристаллической фазы и аморфного гало, по соотношению (4) были
определены относительные величины доли кристаллической фазы \textit{x}($t$).

\begin{figure}
\begin{centering}
\includegraphics[width=10cm]{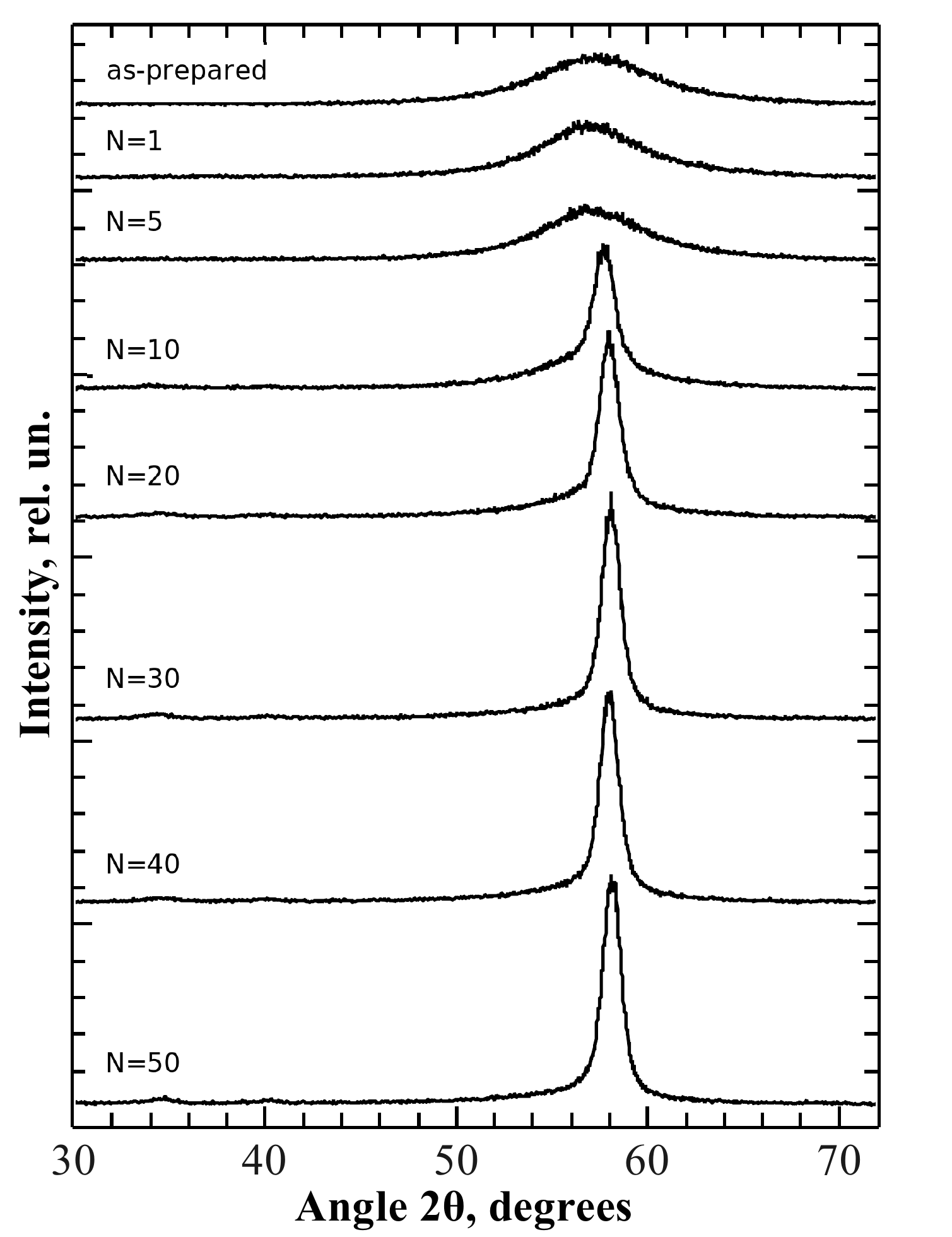}
\par\end{centering}

\caption{Дифрактограммы сплава 5БДСР в исходном состоянии и после облучения
некогерентным оптическим излучением с подводимой энергией $E_{e}$~=~917~Дж
разным количеством импульсов \textit{N}}

\end{figure}

Мы считаем, что световую импульсную обработку можно свести к изотермическому
отжигу и использовать модель КДМА для анализа кинетики кристаллизации.
В самом деле, поскольку подводимая к лампе-вспышке энергия была одинаковой
для всех импульсов, мы предполагаем, что температура центральной части
образца не превышала некоторого максимального значения, поскольку
остальные условия эксперимента были одинаковыми (например, положение
образца относительно лампы, размер образца и т.п.). При импульсном
облучении изменение температуры образца \textit{T} за время \textit{t}
естественно должно зависеть от времени облучения. Правильно при этом
учитывать тепловую инерционность образца, но мы упрощено считаем что
\textit{T}(\textit{t}) повторяет форму импульса, и не учитываем время
нагрева и остывания образца. В этом случае, мы вводим понятие условного
времени отжига $t=N\times\tau$, где \textit{N} \textemdash{} количество
импульсов, а $\tau$ \textemdash{} длительность электромагнитного
импульса лампы-вспышки равная около 500~мкс.

На рис. 6 представлена зависимость от $\ln t$, из которой были найдены
значения \textit{n} и $\ln K$, и во вкладке приведена зависимость
роста кристаллической фазы от условного времени отжига \textit{x}($t$).

\begin{figure}
\begin{centering}
\includegraphics[scale=0.4]{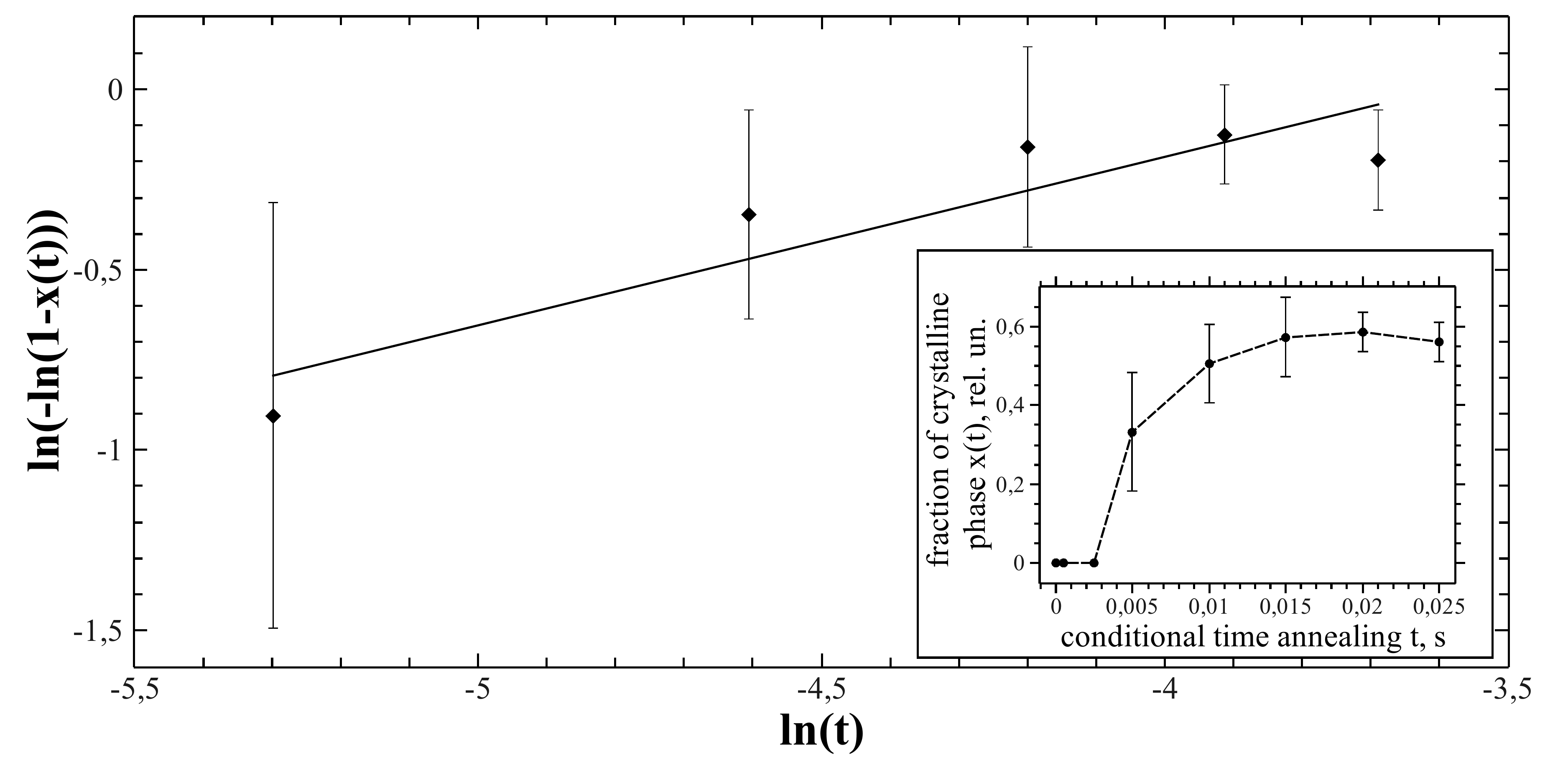}
\par\end{centering}

\caption{Зависимость роста кристаллической фазы в аморфной матрице, представленная
моделью кинетики кристаллизации КДМА}

\end{figure}

Найденные из графика Аврами (рис.~6) значения \textit{n} и \textit{K}
приблизительно равны соответственно 0,5 и 2~эВ. Необычно низкое значение
показателя Аврами $n\approx0,5\pm0,2$ совпадает со значением, представленным
в работах \cite{FINE006,FINE007} в которых кинетику кристаллизации
сплава Fe$_{73,5}$Cu$_{1}$Nb$_{3}$Si$_{13,5}$B$_{9}$, при термическом
отжиге в вакууме, исследовали методами рентгеновской дифракции и Мёссбауэровской
спектроскопии.

Столь низкое значение \textit{n} связано с тем, что кристаллизационные
зародыши возникают на границах медных кластеров размером 3-5~нм,
а не в любом месте аморфной фазы \cite{HONO007,HONO001}. Такие кластеры
меди начинают появляться в объеме аморфной фазы в результате спинодального
распада, что и обуславливает образование их при температурах ниже,
чем температура кристаллизации 1-й стадии. Наличие медных кластеров
существенно понижает энергию, необходимую для образования $\alpha$-Fe(Si)
\cite{CRYS002}. Вероятно именно это обстоятельство является причиной
образования нанокристаллической фазы только после нескольких импульсов
(свыше 5-ти, см. рис.~5 и 6 вкладка). Подводимой энергии $E_{e}$~=~917~Дж
для генерации светового импульса вероятно не хватает для кристаллизации
нанокристаллов $\alpha$-Fe(Si) сразу за один импульс. За несколько
импульсов по объему аморфной фазы успевают сформироваться множество
медных кластеров, которые понижают активационную энергию, необходимую
для образования фазы $\alpha$-Fe(Si). Увеличение же подводимой энергии
до 1310~Дж приводит к \guillemotleft{}двойной\guillemotright{} кристаллизации:
вместе с фазой $\alpha$-Fe(Si) появляется H-фаза. На рис.~7 показаны
дифрактограммы сплава 5БДСР до и после облучения лампой-вспышкой с
подводимой энергией 1310~Дж. Из них видно, что с ростом количества
импульсов \textit{N} соотношение между этими фазами меняется весьма
незначительно.

\begin{figure}
\begin{centering}
\includegraphics[width=10cm]{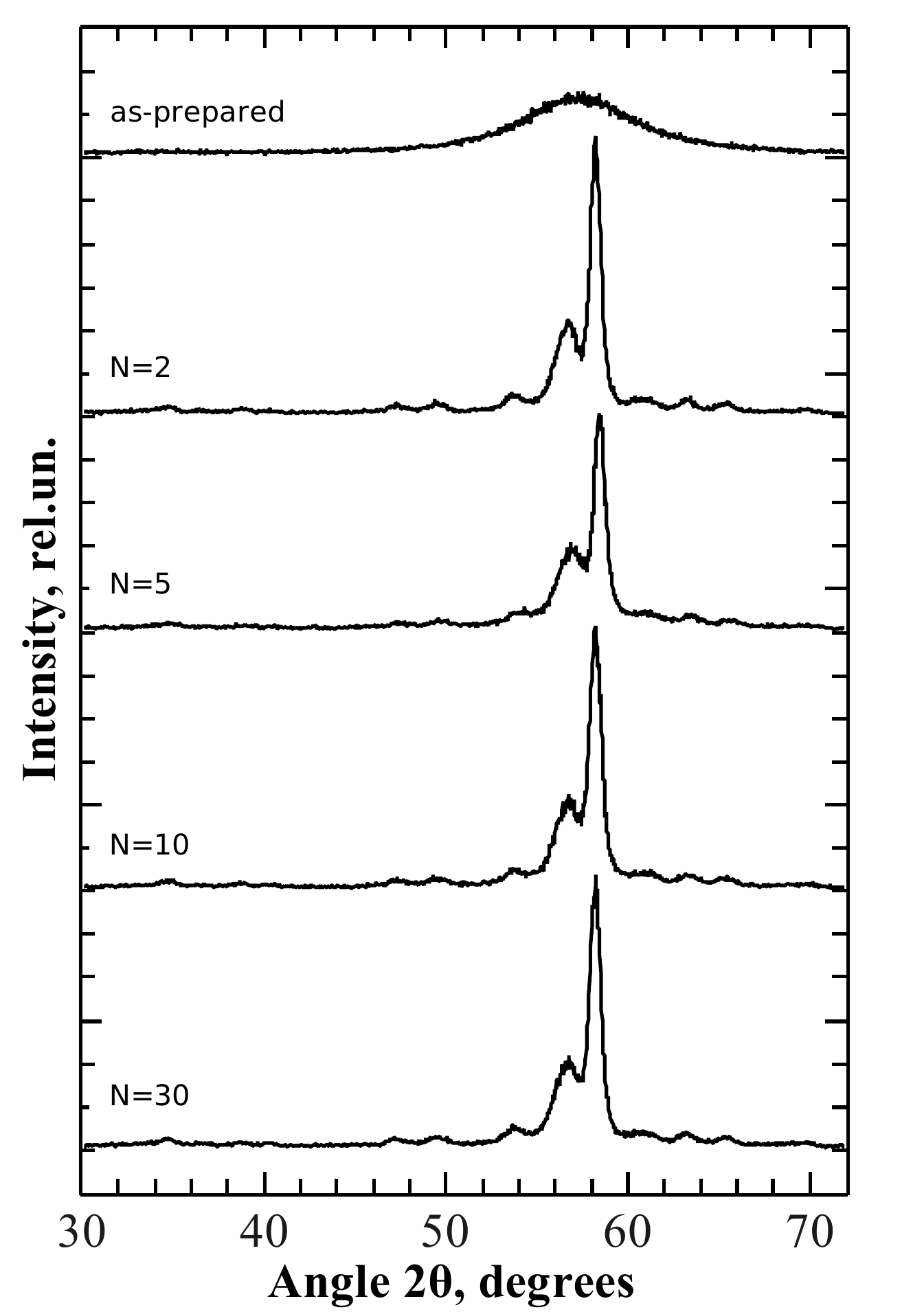}
\par\end{centering}

\caption{Дифрактограммы сплава 5БДСР в исходном состоянии и после облучения
некогерентным оптическим излучением с подводимой энергией $E_{e}$~=~1310~Дж
и разным количеством импульсов \textit{N}}

\end{figure}

Поскольку определяющим фактором магнитомягких свойств является размер
кристаллитов в аморфной матрице, нами по формуле Селякова-Шерера \cite{XRAY001,GUS001}
были определены размеры кристаллитов в зависимости от условного времени
отжига. Для определения физического уширения в качестве эталона использовалась
дифрактограмма образца, облученного световым импульсом с подводимой
энергией $E_{e}$~=~1529~Дж. Как это уже отмечалось ранее, такой
образец мы считали полностью кристаллизованным, с поликристаллической
структурой, в которой размер зерен превышал 150~нм. Из рис.~8 видно,
что наблюдается рост размера зерен \textit{D}~нм (\textpm{}2-3~нм)
при увеличении условного времени отжига \textit{t}.

\begin{figure}
\begin{centering}
\includegraphics[scale=0.4]{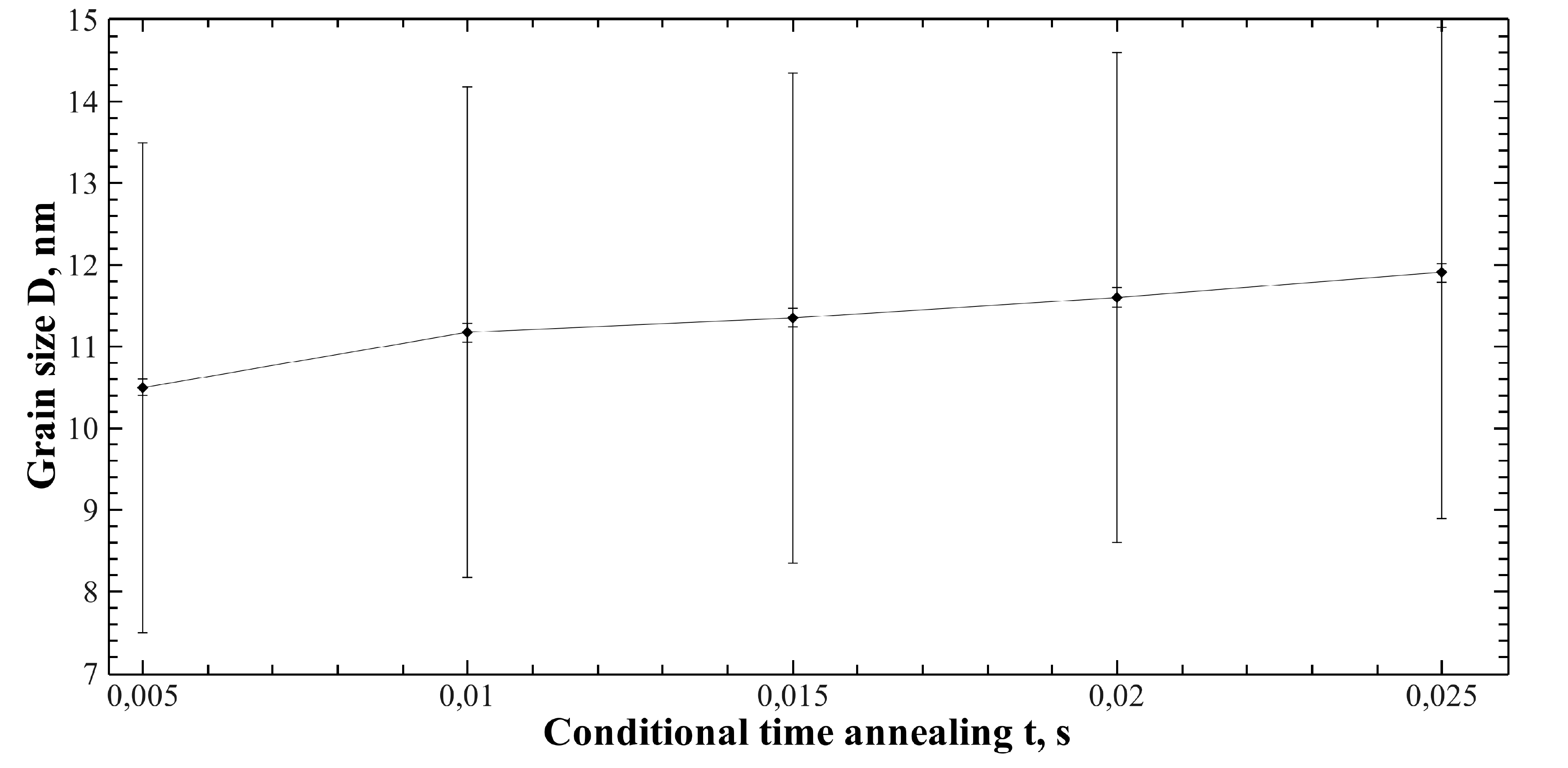}
\par\end{centering}

\caption{Зависимость размера кристаллических зерен \textit{D} от условного
времени отжига t}

\end{figure}

Интересно отметить, что размер кристаллитов увеличивается даже тогда,
когда рост доли кристаллической фазы \textit{x}(\textit{t}) практически
прекратился. Вероятнее всего это связано с укрупнением мелких кристаллитов
вследствие их объединения, и при увеличении условного времени отжига
следует ожидать насыщения размеров кристаллических зерен \textit{D}.

Несмотря на то что размер зерен определен с точностью 20-30\%, мы
считаем, что приведенный на рис.~8 рост обоснованным, поскольку точность
определения физического уширения из дифрактограмм около 0,1-1\% (малый
интервал ошибок на рис.~8). В то же время точность определения \textit{D}
в большей степени зависит от выбора эталона для определения геометрического
уширения и учета дополнительных вкладов (дублетность линии $K_{\alpha}$,
возможные вклады от микроискажений решетки и химической неоднородности)
в физическое уширение, которые в настоящей работе не проводились.

\section{Выводы}

Представленные результаты показали, что в результате импульсного облучения
некогерентным электромагнитным излучением в диапазоне от ИК до УФ,
генерируемого газоразрядной лампой-вспышкой, аморфный сплав 5БДСР
системы Fe-Cu-Nb-Si-B кристаллизуется с образованием ряда промежуточных
структур, состояние которых зависит от подводимой к лампе электрической
энергии. Найдены величины подводимой энергии при которых облучаемый
сплав:
\begin{enumerate}
\item остается аморфным, но меняется ближний порядок;
\item становится нанокристаллическим, с межзеренной границей, состоящей
из аморфной фазы либо из мелкодисперсной H-фазы;
\item становится поликристаллическим, с крупным размером зерен (более 150~нм).
\end{enumerate}
Кинетика фазового перехода в области нанокристаллизации была исследована
с использованием рентгеновской дифракции. Эти исследования показали,
что формирование нанокристаллов при импульсном облучении идет по механизму,
аналогичному для системы Fe-Cu-Nb-Si-B, но при изотермическом отжиге
из аморфного состояния. Кристаллизация происходит только при определенной
подводимой энергии и начинается при количестве импульсов свыше \textit{N}~=~5,
что можно рассматривать как условное время возникновения кластеров
меди и кристаллизационных зародышей фазы $\alpha$-Fe(Si) на границах
с этими кластерами. Малое значение показателя Аврами $n\approx0,5$
показывает, что формирование зародышей происходит до роста кристаллических
зерен $\alpha$-Fe(Si) и их рост сильно ограничен (сильнее, чем при
диффузионно-барьерном росте).

\end{document}